\documentclass[showpacs,preprintnumbers,amsmath,amssymb,a4paper,superscriptaddress]{revtex4}
\usepackage{amsmath}
\usepackage{graphicx}
\usepackage{dcolumn}
\usepackage{bm}
\usepackage{booktabs}
\usepackage{calc}
\usepackage{multirow}
\usepackage{mathtools}
\usepackage{color}

\DeclareMathAlphabet{\mathpzc}{OT1}{pzc}{m}{it}

\newcommand {\beq} {\begin{eqnarray}}
\newcommand {\eeqn} [1] {\label{#1} \end{eqnarray}}

\newcommand{\rel}[1]{\rho\,}
\newcommand{\ra}[1]{\hat{\mathbf{r}}\,}
\newcommand{\ri}[1]{\mathbf{R}\,}
\newcommand{\ma}[1]{m\,}
\newcommand{\mi}[1]{M\,}
\newcommand{\xa}[1]{x}
\newcommand{\ya}[1]{y\,}
\newcommand{\za}[1]{z\,}
\newcommand{\X}[1]{X\,}
\newcommand{\Y}[1]{Y\,}
\newcommand{\Z}[1]{Z\,}

\begin{document}

\title{Wave-like amplification of near-threshold two-particle reactions: from muon-catalyzed fusion to $\Lambda\bar{\Lambda}$ production at $e^-e^+$ annihilation}
\author{Vladimir S.Melezhik}
\email[]{melezhik@theor.jinr.ru}
\affiliation{Bogoliubov Laboratory of Theoretical Physics, Joint Institute for Nuclear Research, 6 Joliot-Curie St., Dubna, Moscow Region 141980, Russian Federation}
\affiliation{Dubna State University, 19 Universitetskaya St., Dubna, Moscow Region 141982, Russian Federation}
\date{\today}
\begin{abstract}\label{txt:abstract}
A simple model is proposed to explain the recently found wave-like enhancement of the $\Lambda\bar{\Lambda}$ pair production near the threshold at the $e^-e^+$ annihilation, which allows extracting model-independent scattering parameters and spectral information for the $\Lambda\bar{\Lambda}$ pair from the oscillatory nature of the measured cross section. In particular, it predicts a single bound state of $\Lambda\bar{\Lambda}$ with a binding energy of $\varepsilon_{\Lambda\bar{\Lambda}}=(36\pm5)$MeV. The model is a generalization of the formulas obtained in our earlier work [1] to explain the effect of wave-like amplification found in it near the threshold of fusion reactions screened by a muon or electron. The analysis allows us to conclude that the effect of wave-like amplification is an integral feature of any two-particle near-threshold reaction. In this regard, it seems promising to investigate, within the framework of our model, the oscillatory nature of the electromagnetic form factors of hyperons and nucleons extracted in experiments on $e^- e^+$ annihilation. A natural further development of the model could be its generalization to processes of producing various hadron pairs in $e^-e^+$ annihilation.
\end{abstract}

\maketitle

\section{Introduction}

In our work \cite{Mel}, based on the two-channel potential approach \cite{Bogd1,Bogd2} for calculating  the partial cross sections $\sigma_J(E)$ of fusion reactions
\begin{align}
t + d \rightarrow ^{4}He + n + 17.6 \,\textrm{MeV}\,\,\\
d + d \rightarrow ^{3}He +n + 3.27\, \textrm{MeV}\,\, (t + p + 4.03\, \textrm{MeV}) \,\,,
\end{align}
the low-energy limit ($v=\sqrt{2E/M}\rightarrow 0$) was obtained for the cross sections
\begin{align}
\sigma_J(E)= A_J v^{2J-1} \mid f_J(E)\mid^{-2}\,,
\end{align}
where $f_J(E)=\mid f_J(E)\mid\exp(-i\delta_J(E))$ is the Jost function \cite{Taylor}
\begin{align}
(kr)^J\mid f_J(E)\mid^{-1} \,\, &\mathop{\longleftarrow}\limits_{0\leftarrow kr}\,\,\psi_J(kr) \mathop{\longrightarrow}\limits_{kr\rightarrow
\infty}\, \frac{\sin(kr-\pi J/2 +\delta_J(E))}{kr}
\end{align}
and $\psi_J(kr)$ is the radial wave function describing the process of pair collision in the input channel of the reaction $X+Y\rightarrow X'+Y'$ in the partial wave $J$ at $E\geq 0$, $k=\sqrt{2M E}/\hbar$.
From definition (4) it follows that for the s-wave ($J=0$) the relation $\mid f_0(E)\mid^{-2}=\mid\psi_0(kr)\mid^2_{r=0}$ is satisfied. In this case, formula (3) coincides with the well-known low-energy approximation $\sigma_0(E)=A_0/v\mid \psi_0(r=0)\mid^{2}$ \cite{Deser,Leng}, which, in the case of a purely Coulomb interaction $z_Xz_Ye^2/r$ between particles in the input channel $X+Y$, goes over to the formula
\begin{align}
\sigma_0(E)=\frac{A_0}{v}C_0^2(v)\,,
\end{align}
widely used for the analysis in the low-energy limit of experimental data on fusion reactions of light nuclei (see, for example, \cite{Leng}). Here $C_0^2(v)=2\pi\eta/(\exp(2\pi\eta)-1)$ is the Gamow-Sommerfeld factor, $\eta=z_Xz_Y\alpha c/v$, $\alpha=1/137$ is the fine structure constant, $c$ is the speed of light in vacuum, and $z_X$ and $z_Y$ are the charges of the colliding particles. In works \cite{Bogd1,Bogd2}, to analyze reactions (1) and (2), the Coulomb interaction in the $X+Y$ channel at small distances between the interacting nuclei was replaced by a model nuclear potential. The constants $A_J=\lim_{E\rightarrow\infty} \sigma_J(E) v^{1-2J}\mid f_J(E)\mid^2$ of reactions (1) ($J=0$) and (2) ($J=1$) obtained in this approach were used to calculate the rates of fusion reactions in mesic molecules $dt\mu$ \cite{Bogd1} and $dd\mu$ \cite{Bogd2}, which were subsequently confirmed experimentally \cite{Ponomarev,Petitjean}.

\begin{figure}
\centering\includegraphics[scale=0.33]{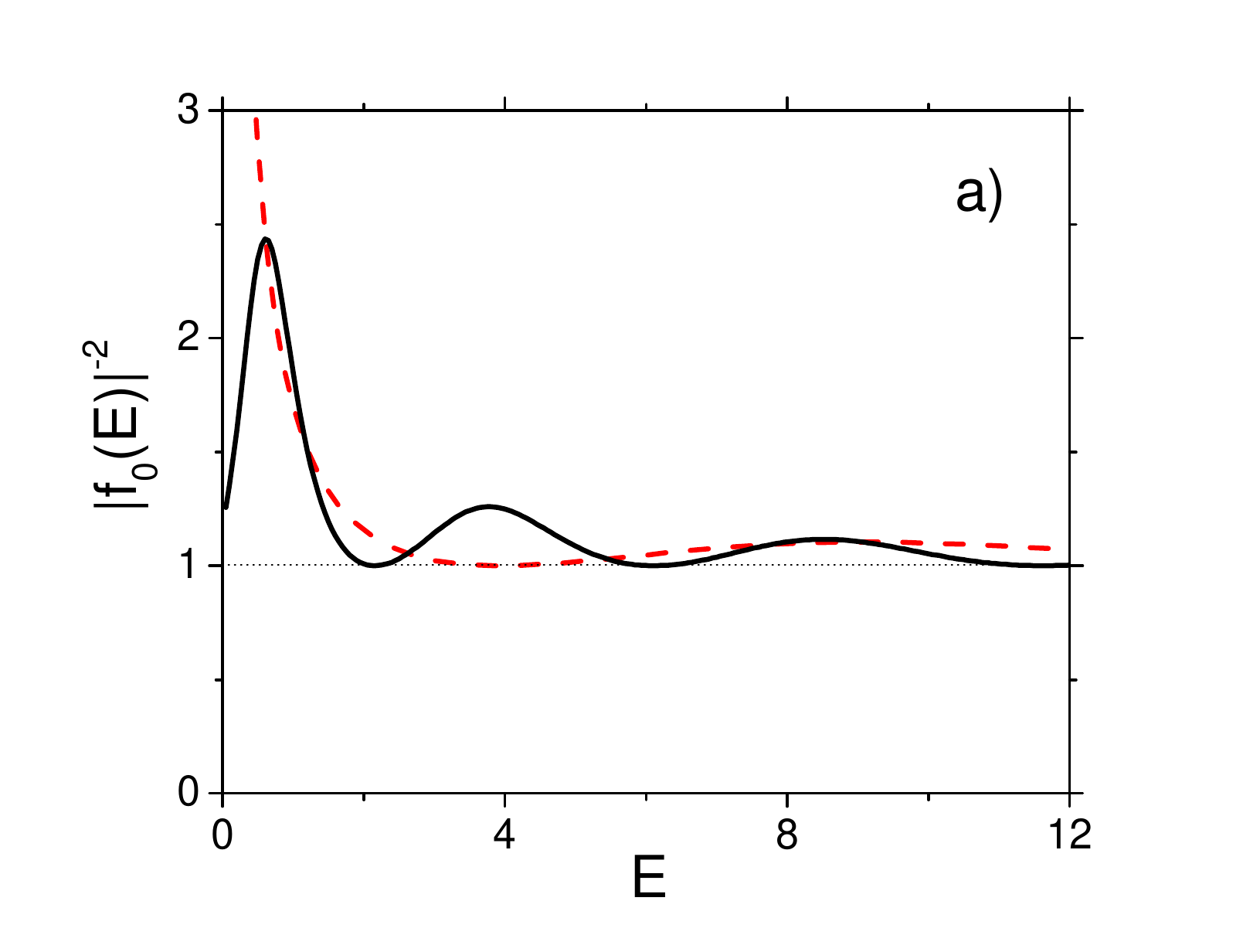}
\vspace{-0.5cm}
\centering\includegraphics[scale=0.33]{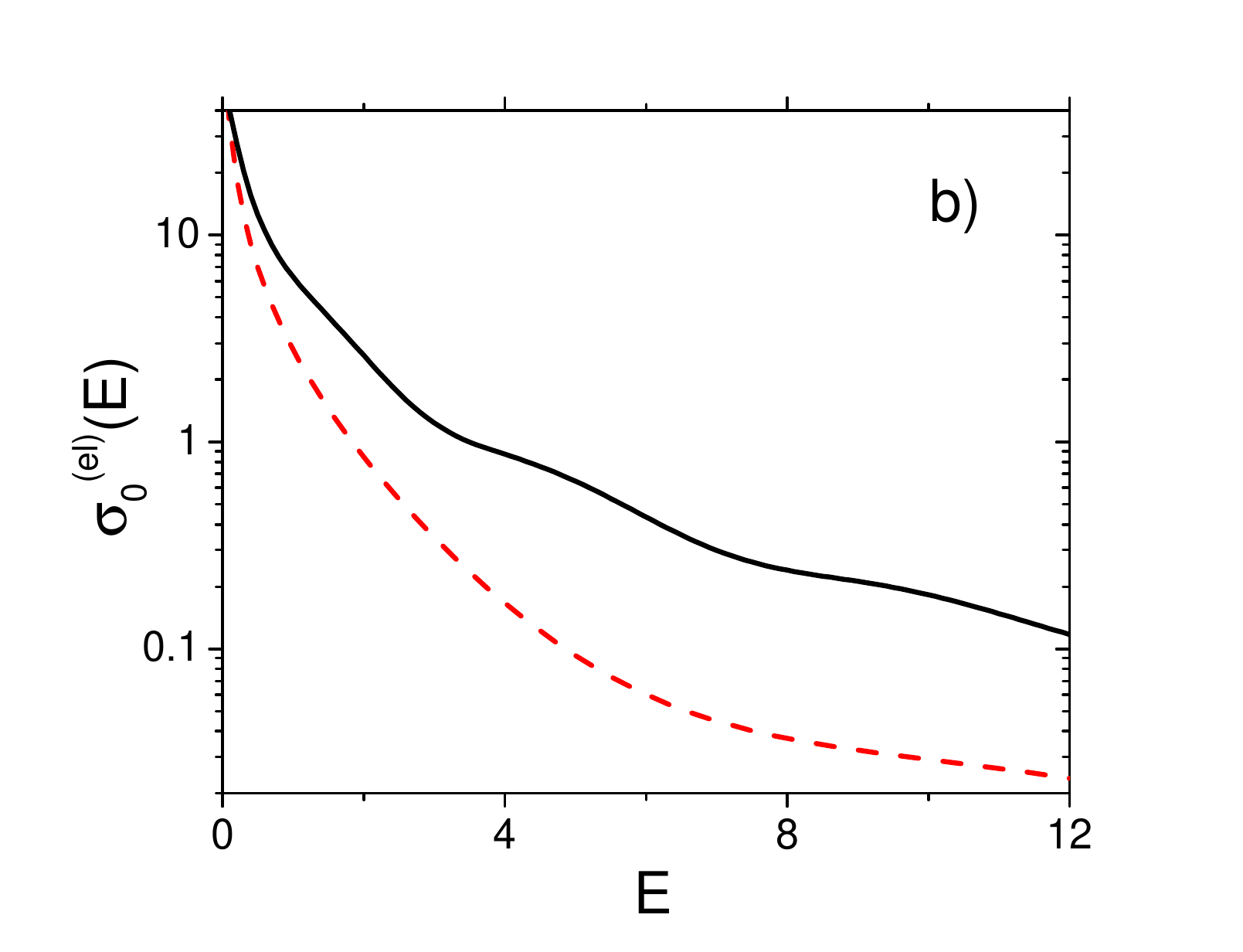}
\vspace{-0.4cm}
\centering\includegraphics[scale=0.33]{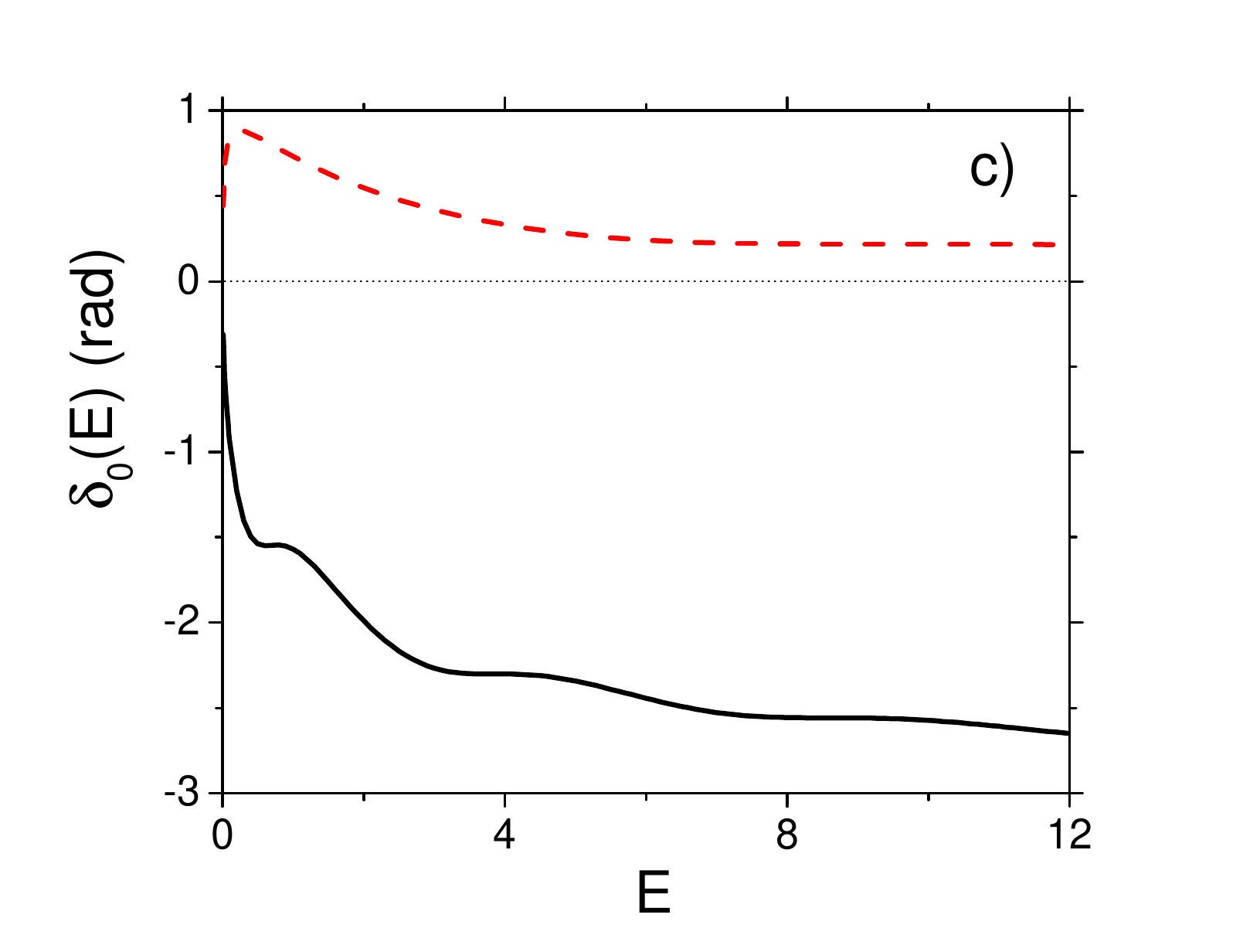}
\caption{Dependence of the function $\mid f_0(E)\mid^{-2}$ (a), the s-wave elastic scattering cross section $\sigma_0^{(el)}(E)$ (b) and the corresponding scattering phase shift $\delta_0(E)$ (c) on the energy $E>0$ for a spherically symmetric potential wells $\hbar=M=V_0=1$. The dashed curves correspond to a potential well in which there are no bound states ($R_0=1, \nu=0$). In this case, the maxima of the crests of the function $\mid f_0(E)\mid^{-2}$ closest to zero and the next in energy are located at the points $E_0=0.23$ and $E_1=10.1$, respectively. When the well expands to $R_0=2.5$, a level with energy $\varepsilon_{\nu=1} =-0.540$ appears in it, and the positions of the first three maxima of the function $\mid f_0(E)\mid^{-2}$ closest to zero are at the points $E_1=0.78, E_2=3.93$ and $E_3=8.66$. This case is represented by solid curves.}
\end{figure}
As follows from the above, the energy dependence of the quantity $\sigma_0(E)v$ for the reaction $X+Y\rightarrow X'+Y'$ in the low-energy limit is determined by the Jost function of the pair $X+Y$ (by the Gamow factor $C_0^2(v)$ for charged particles $X$ and $Y$). The fact that a large volume of experimental data on the cross sections of various fusion reactions is perfectly described by formula (5) in the experimentally achievable region of $E\gtrsim(5-10)$ keV stimulated the use of a smooth exponential energy dependence of the factor $C_0^2(E)$ for extrapolating this quantity to the region of the ``Gamow window'' at low energies (see, for example, \cite{Leng,Angulo}). However, when calculating the functions $\mid f_0(E)\mid^{-2}$ in the adiabatic representation of the three-body problem, which determine the low-energy behavior of $\sigma_0(E)v$ for three-body reactions
\begin{align}
p\mu + p \rightarrow d + e^+ +\nu_e + \mu^-\,\,\\
D + d \rightarrow ^{3}He + n + e^- (t + p + e^-) \,\,,
\end{align}
their wave-like dependence on $E$ was found, with an increase in the limit $E\rightarrow 0$ \cite{Mel}. This effect was later confirmed independently by calculating $\sigma_0(E)v$ in the Faddeev approach \cite{Kvits} for muon-catalyzed reactions of type (6). It turned out that it is already present in the simplest model \cite{Mel}, approximating the interaction in the $X+Y$ channel with a spherically symmetric rectangular potential well with depth $V_0$ and width $R_0$. Indeed, in this case, the function $\mid f_0(E)\mid^{-2}$ is described by the formula \cite{Mel}(see Fig.1a)
\begin{align}
\mid f_0(E)\mid^{-2} = \frac{E+V_0}{E+V_0\cos^2(qR_0)}\,\,,
\end{align}
where $q=\sqrt{2M(E+V_0)}/\hbar$, $M$ is the reduced mass of the pair $X+Y$. That is, the energy dependence of the quantity $\sigma_0(E)v=A_0 \mid f_0(E)\mid^{-2}$ has an oscillatory nature with an oscillation amplitude from $A_0$ to
\begin{align}
\sigma_0(E_n)v_n = A_0(1+\frac{V_0}{E_n})
\end{align}
as we move away from the threshold $E=0$, due to the periodic dependence on the energy of the denominator in formula (8). The positions of the maxima $E_n$ (9) are determined from the condition $\cos(qr_0)=0$ by the formula
\begin{align}
E_n=\frac{1}{2M}(\frac{\pi\hbar(2n+1)}{2R_0})^2-V_0\,,
\end{align}
where $ n $ takes the values $n= \nu, \nu =1, ...$, starting with $\nu$, equal to the number of bound states in the potential well. Moreover, if the maximum closest to the threshold is located sufficiently close to it ($E_{\nu} \rightarrow 0$), then in the elastic scattering cross section
$$
\sigma_0^{(el)}(E)=\frac{4\pi}{k^2}\sin \delta_0(E)=\frac{const}{E+E_{\nu}}\,
$$
an s-wave resonance is observed at the virtual level (Fig.1b), known in the literature as the Bethe-Peierls resonance, where the dependence of the scattering phase $\delta_0(E)$ on $E$ has a resonant character (Fig.1c). In this case, if the closest to the threshold $E=0$ crest of the function $\mid f_0(E)\mid^{-2}$ corresponds to the resonant behavior of the phase shift $\delta_0(E)$ in the vicinity of the point $E=E_{\nu}$ and a near-threshold resonance in the cross section $\sigma_0^{(el)}(E)$, then the subsequent crests at the points $E=E_{\nu+1}$ and $E_{\nu+2}$ correspond to plateaus in the energy dependencies of the functions $\delta_0(E)$ and $\sigma_0^{(el)}(E)$ in the vicinity of these points.

As the energy moves away from the resonance at the point $E=E_{\nu}$, the function $\mid f_0(E)\mid^{-2}$ approaches unity in a wave-like manner, with decreasing oscillation amplitude (9) and increasing ``wavelength'' (the distance between wave crests)
\begin{align}
E_{n+1}-E_n=\frac{(n+1)}{M}(\frac{\pi\hbar}{R_0})^2\,,
\end{align}
which is determined by the interaction radius $R_0$ between particles $X$ and $Y$ and their reduced mass $M$. It follows that by determining the positions of $E_{\nu}, E_{\nu+1}$, and $E_{\nu+2}$ of the three crests closest to the threshold for the experimentally measured quantity $\sigma_0(E)v$, one can estimate the number of bound states $\nu$ for the pair $X+Y$ using the simple formula
\begin{align}
\frac{E_{\nu+2}-E_{\nu+1}}{E_{\nu+1}-E_{\nu}}=\frac{\nu+2}{\nu+1}\,.
\end{align}
Further, using formula (11), one can also estimate the magnitude of the interaction radius $R_0$ from the experimentally measured difference $E_{\nu+1} - E_{\nu}$ and the depth $V_0$ of the interaction potential from the magnitude of the measured $\sigma(E_n)v_n$ at these points (9).

In this work, we show that a simple model based on formulas (3,4),(8-12) also satisfactorily describes the recently observed \cite{Bes1,Bes2} wave-like enhancement of the reaction
\begin{align}
e^- + e^+ \rightarrow \Lambda + \bar{\Lambda}\,
\end{align}
near the $\Lambda\bar{\Lambda}$ pair production threshold and allows us to extract model-independent scattering parameters, the scattering length $a_s$ and the effective radius $r_0$, of the $\Lambda\bar{\Lambda}$ pair and information about its spectrum.

\section{Theoretical model for $e^- + e^+ \rightarrow \Lambda + \bar{\Lambda}$}

The $\Lambda\bar{\Lambda}$ pairs are produced mainly in the s-state ($J=0$), with a fixed isotopic spin ($I=0$) \cite{Bes1}, and the Coulomb interaction between $\Lambda$ and $\overline{\Lambda}$ is absent. Moreover, the possibility of the $\Lambda\bar{\Lambda}$ pair annihilation into mesons, has only a minor effect on the reaction cross section (13) \cite{Mil3,Mil4}. Therefore, it is natural to use formula (3) for $J=0$, obtained for reaction (1) in [1], to describe the reaction
\begin{align}
\Lambda + \bar{\Lambda}\rightarrow e^- + e^+\,\,,
\end{align}
which is the inverse of (13). Formula (3) was obtained in \cite{Mel} for fusion reaction (1) within the framework of the two-coupled-channel model \cite{Bogd1}, in which the interaction of $d$ and $t$ was described by a generalized optical potential, the structure of which was established from the consideration of the problem of two coupled channels $d+t$ and $^4He +n$: the anti-Hermitian part had a separable form, and the Hermitian part was approximated by a local and energy-independent potential. It should also be noted that when deriving formula (3) \cite{Bogd1}, the square of the matrix element $\langle dt\mid \hat{V}_{12}(E+\Delta E)\mid n^4He \rangle$, which couples the $d+t$ and $^4He +n$ channels in reaction (1), was included in the constant $A_0$, since the distance between their thresholds $\Delta E=17.6$ MeV significantly exceeded the considered region of collision energies $E\lesssim 100$ keV between $d$ and $t$. However, for reactions (13) and (14), the energy dependence of this matrix element, which in this case is determined by the dipole form factor $F_D(E)$, is significant since the distance $\Delta E=2m_{\Lambda}c^2=2.231$ GeV between the thresholds of the $\Lambda+\bar{\Lambda}$ and $e^- + e^+$ channels and the value $9$ MeV$\lesssim E \lesssim 700$ MeV of the considered region of kinetic energies of the $\Lambda\bar{\Lambda}$ pair become comparable. To account for this dependence, we use $F_D(E)$ accepted in describing reactions of type (13) \cite{Halzen,Mil2,Mil3,Mil4}
\begin{align}
A_0 \rightarrow A_0F_D^2(E)=A_0(1-s/\Delta^2)^{-4}\,,
\end{align}
where $s=4(m_{\Lambda}^2c^4+p_{\Lambda}^2c^2)=4(m_{\Lambda}^2c^4+m_{\Lambda}c^2E)=4(m_e^2c^4+p_e^2c^2)$ is the Mandelstam parameter, $E$ is the kinetic energy of the $\Lambda\bar{\Lambda}$ pair, $p_{\Lambda}$ and $p_e$ are the three-dimensional momenta of the electron and $\Lambda$- hyperon, respectively, $\Delta$ is the fitting parameter. Then the cross section of reaction (14) is described by the formula
\begin{align}
\sigma_0'(E)=A_0\frac{F_D^2(E)}{v}\mid f_0(E)\mid^2\,.
\end{align}
To return to the reaction (13) of interest to us, we use the principle of detailed balance
\begin{align}
\sigma_0(E)=\frac{p_{\Lambda}^2}{p_{e}^2}\sigma_0'(E)=\frac{p_{\Lambda}^2}{p_{e}^2v}F_D^2(E)A_0\mid f_0(E)\mid^2\,.
\end{align}
Using the definition of the Mandelstam parameter $s$ and neglecting the rest energy $2m_ec^2=1.02$ MeV of the $e^-e^+$ pair compared to their total energy $p_e^2/m_e \sim 2m_{\Lambda}c^2=2.231$ GeV, we obtain
\begin{align}
\frac{p_{\Lambda}^2}{p_e^2v}=\frac{s-4m_{\Lambda}^2c^4}{sv}=\frac{\sqrt{m_{\Lambda}E}}{m_{\Lambda}c^2+E}\,.
\end{align}
Thus, the cross section of reaction (13) is expressed in terms of the modulus of the Jost function $\mid f_0(E)\mid$ of the relative motion of $\Lambda$ and $\bar{\Lambda}$ and the dipole form factor $F_D(E)$ as follows
\begin{align}
\sigma_0(E)=\frac{\sqrt{m_{\Lambda}E}}{m_{\Lambda}c^2+E}F_D^2(E)A_0\mid f_0(E)\mid^{-2}\,.
\end{align}

\section{Results and discussions}

By varying the free parameter $\Delta$ in the dipole form factor $F_D(E)$ (15), we achieved that the experimental cross sections $\sigma_0^{(exp)}(E)$, ``renormalized'' by the factor $\sqrt{m_{\Lambda}}F_D^2/(m_{\Lambda}c^2+E)$ from (19)
\begin{align}
 \sigma_0^{(exp)}(E)\frac{m_{\Lambda}c^2+E}{F_D^2(E)\sqrt{m_{\Lambda}E}}=\bar{\sigma}_0(E)v = A_0\mid f_0(E)\mid^{-2}\,\,,
\end{align}
qualitatively reproduced the wave-like behavior of the function $A_0\mid f_0(E)\mid^{-2}$ for a rectangular potential well (8)(Fig.1a) with an approximately constant depth $A_0$ of its troughs. The points $\bar{\sigma}_0(E)v$ calculated by formula (20) from experimental cross sections $\sigma_0^{(exp)}(E)$ are shown in Fig.2 for $\Delta=1.7$ GeV. Here the curve $\mid f_0(E)\mid^{-2}$ calculated using formula (8) for a potential well with $V_0=58.3$ MeV and $R_0=3.10$ fm is also presented. The parameters of the potential well $V_0$ and $R_0$ and the constant $A_0=6.52\times 10^{-24}$cm$^3$/sec were found from the condition of the best approximation of the curve $A_0\mid f_0(E)\mid^{-2}$ to the points $\bar{\sigma}_0(E)v$.
Figure 3 shows the experimental $\sigma_0^{(exp)}(E)$ and theoretical $\sigma_0(E)$ cross sections of reaction (13). The curve $\sigma_0(E)$ was calculated by formula (19) with the parameters $V_0$, $R_0$ and $A_0$ fixed for $\Delta=1.7$ GeV. For comparison, here are also given the theoretical dependencies of $\sigma_0(E)$ on $E$, calculated for $\Delta=1.8$ GeV and $1.6$ GeV, allowed by the error corridor $\pm\Delta\sigma(E)$ of the experimental cross section $\sigma_0^{(exp)}(E)$. It should be noted that outside the interval $1.6$GeV$\leq\Delta\leq 1.8$GeV, we were unable to find such a set of parameters $V_0, R_0$ and $A_0$ that the theoretical curve (19) satisfactory describes all the experimental points shown in Fig.3, within the experimental errors.

The wave-like behavior of the $\bar{\sigma}_0(E)v$ curve constructed over the experimental points $\sigma_0^{(exp)}(E)$ as a function of $E$ (see Fig.2) allows us to draw a positive conclusion about the presence of one bound state in the $\Lambda \overline{\Lambda}$ pair. Indeed, using formula (12) for the three crests of $\bar{\sigma}_0(E)v$ closest to the pair production threshold $\Lambda \overline{\Lambda}$ at $E_{\nu}=9.13$ MeV, $E_{\nu+1}=153.6$ MeV, and $E_{\nu+2}=348.7$ MeV, we obtain
\begin{align}
\frac{\nu+2}{\nu+1}=\frac{E_{\nu+2}-E_{\nu+1}}{E_{\nu+1}-E_{\nu}}\simeq 1.4\,,
\end{align}
from which the estimate $\nu \simeq 1$ follows. Direct calculation confirms the presence of one bound state with the energy $\varepsilon_{\nu=1}=-36$ MeV in the rectangular potential well with $V_0=58.3$ MeV and $R_0=3.1$ fm, which approximates the Hermitian part of the generalised optical potential describing the $\Lambda\bar{\Lambda}$ interaction in our approach \cite{Mel,Bogd1}. Moreover, the variation of the parameter $\Delta$ in $F_D(E)$ (15) within the range of $(1.6-1.8)$ GeV, allowed by the experimental errors [11], gives an error of an order of $\Delta\varepsilon_{\nu=1}=\pm 5$ MeV in determining the position of the level $\varepsilon_{\nu=1}$ in the frame of our model.

\begin{figure}
\centering\includegraphics[scale=0.36]{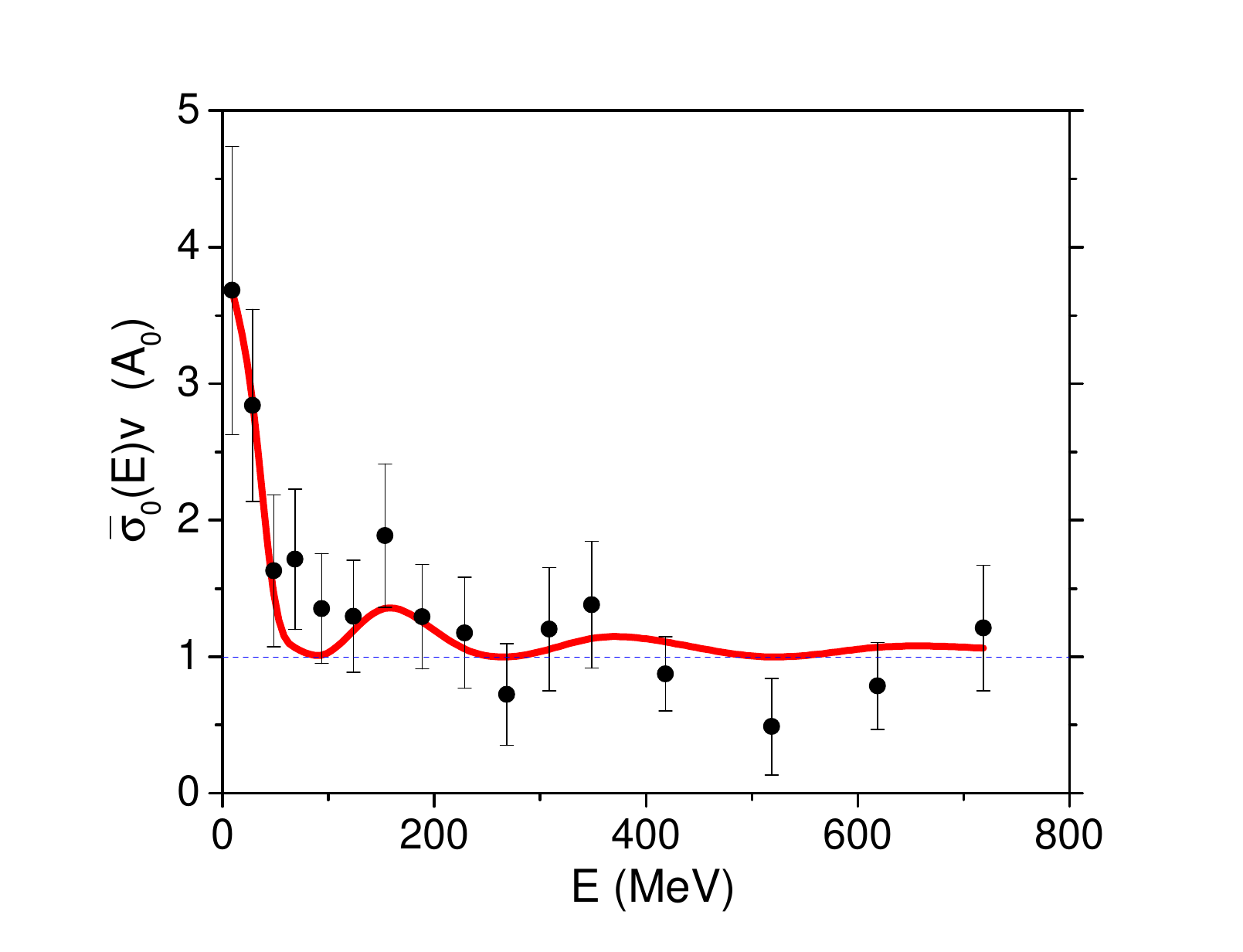}
\caption{Points $\bar{\sigma}_0(E)v$ recalculated by formula (20) from the experimental cross sections $\sigma_0^{(exp)}(E)$ of reaction (13) \cite{Bes1}, and the function $\mid f_0(E)\mid^{-2}$ calculated by formula (8) for a potential well with $V_0=58$ MeV and $R=3.1$ fm. The values $\bar{\sigma}_0(E)v$ are given in units $A_0=6.5\times 10^{-24}$cm$^3$/sec.}
\end{figure}
\begin{figure}
\centering\includegraphics[scale=0.40]{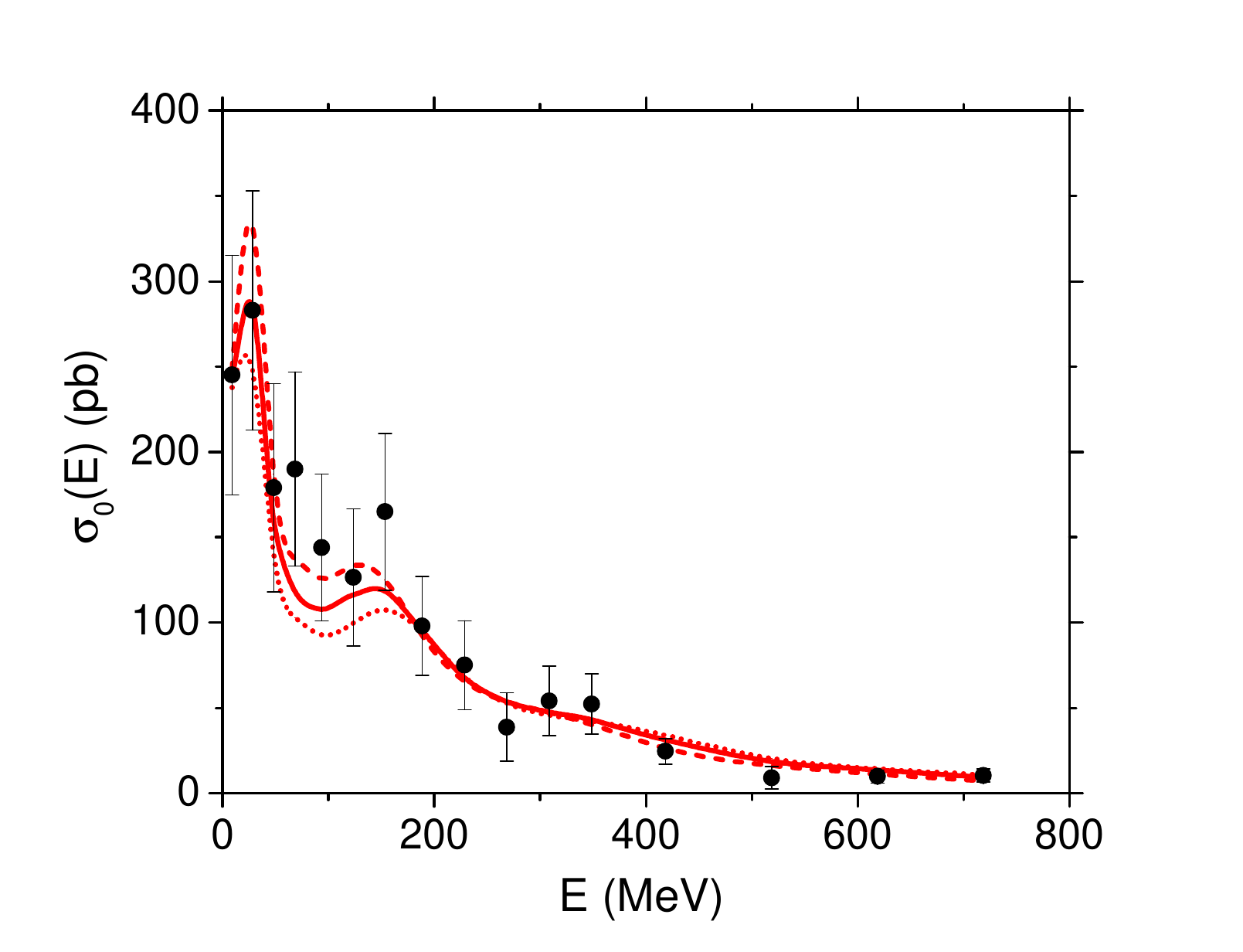}
\caption{Experimental cross sections $\sigma_0^{(exp)}(E)$ of annihilation of $e^-e^+$ into a pair $\Lambda\bar{\Lambda}$ (13) as a function of the $\Lambda\bar{\Lambda}$ kinetic energy together with theoretical curves $\sigma_0(E)$ for these cross sections calculated by formulas (20,16,8) for $\Delta=1.7$ GeV, $A_0=6.5\times 10^{-24}$ cm$^3$/sec, $V_0=58$ MeV,$R_0=3.1$ fm (solid curve); $\Delta=1.8$ GeV, $A_0=2.6\times 10^{-24}$ cm$^3$/sec, $V_0=50$ MeV, $R_0=3.2$ fm (dashed curve) and $\Delta=1.6$ GeV, $A_0=1.5\times 10^{-23}$cm$^3$/sec, $V_0=66$ MeV, $R_0=3.0$ fm (doted curve). The experimental points $\sigma_0^{(exp)}(E)\pm\Delta \sigma(E)$ are taken from \cite{Bes1} (Table 1).}
\end{figure}

Here, we should underline the following. With a rectangular potential well, we only approximated the local Hermitian part of the generalized optical potential (see \cite{Bogd1}) describing the $\Lambda\bar{\Lambda}$ interaction in our model. Therefore, we consider it not as a ``realistic'' potential, but only as a phenomenological potential, which gives an analytical formula (19) convenient for calculating the cross sections of reaction (13) near its threshold. In addition, it also provides analytical expressions for the measurable, model-independent scattering parameters - the scattering length $a_s$ and the effective radius $r_0$. The calculation using these formulas with the values of $V_0$ and $R_0$ fixed above gives a fairly realistic estimate for the scattering length of $\Lambda$ by $\bar{\Lambda}$ $a_s=(2.2\pm0.3)$fm and effective radius $r_0=(0.8\pm0.4)$fm. Indeed, the value $r_0$ obtained in this way is in satisfactory agreement with the range of realistic $N\bar{N}$ potentials of the order of 1 fm or somewhat less (see, for example \cite{El-Bennich,Zhou}).

In conclusion of the discussion, it is worth noting the previously developed theoretical models that take into account the interaction in the final state during the production of hadrons near the threshold at the $e^-e^+$ annihilation, in which the energy dependence of the cross sections of the processes was determined by the energy dependence of the wave function of the relative motion of a nascent pair at small distances (see \cite{Heid1,Heid2,Heid3,Mil1,Mil2,Mil3,Mil4} and references therein). However, the possibility of extracting information about the final-state interaction from the oscillatory nature of the measured cross sections near the threshold has not been considered, to our knowledge. Moreover, in a recent paper \cite{Mil3}, to describe the experimental data \cite{Bes1} on reaction (13) above the $\Lambda\bar{\Lambda}$ production threshold, a narrow rectangular potential well with $V_0=584$ MeV and $R_0=0.45$ fm was specifically chosen so that the second from the threshold crest $E_{\nu+1} \sim 1/R_0^2$ (11) of its s-wave function $\psi_E(r=0)$ of the continuum went far beyond the studied energy region $E\leq 700$ keV $\ll E_{\nu+1} \simeq 6.8$ GeV.
Apparently, therefore, the use of a ``simplified formula'' in \cite{Mil3} for the $|\psi_E(r=0)|^2$-function (Eq.(4) in \cite{Mil3}) describes the increase in the cross section of reaction (13) near the threshold of $\Lambda\bar{\Lambda}$ production, but does not reproduce the oscillations in it.
However, the narrow rectangular well used in \cite{Mil3} also contains a weakly bound state with a binding energy $\approx 30$ MeV close to our evaluation of the binding energy of the $\Lambda\bar{\Lambda}$ pair $\varepsilon_{\lambda\bar{\Lambda}}=(36\pm 5$) MeV.

Finally, it should be noted that the dipole form factor $F_D(E)$ for $\Lambda$-hyperon in the cited work \cite{Mil3} was fixed with a parameter $\Delta=1$ GeV close to the value $\Delta=0.84$ GeV for the proton dipole form factor. In our work, the parameter was fitted and fixed at the level of $\Delta=1.7$ GeV giving the best approximation of the experimental points on reaction (13) with our analytic formula (19).  Such an increase in $\Delta$ for the heavier $\Lambda$-hyperon compared to the proton seems to us quite reasonable. Moreover, such an increase of $\Delta$ in the dipole form factor $F_D(E)$ of the $\Lambda$-hyperon gives a rough, but nevertheless more realistic estimate of the root-mean-square charge radius of the $\Lambda$-hyperon \cite{Halzen}
\begin{align}
r_{rms}=\sqrt{6\frac{dF_{D}(s)}{ds}\mid_{s=0}} = \sqrt{12}\frac{\hbar c}{\Delta} \simeq 0.4 \textrm{fm}\,\,
\end{align}
than the value $r_{rms}=0.81$ fm following from the $\Delta=0.84$ GeV corresponding to the  proton dipole form factor.
In this regard, even our rough estimate $r_{rms}\simeq 0.4$ fm for the $\Lambda$-hyperon is of interest since extracting this parameter from the experimental data on reaction (13) is a non-trivial and relevant task \cite{Bes2}.

\section{Conclusion}

In conclusion, we note the following. In the work, simple formulas (8-12), (15), (19) and (20) were obtained for describing the experimental cross sections of the production of $\Lambda\bar{\Lambda}$ pairs during $e^-e^+$ annihilation near the threshold. In this approach, the observed wave-like enhancement of this reaction near the $\Lambda\bar{\Lambda}$ production threshold \cite{Bes1,Bes2} is naturally explained by the influence of interparticle interaction in the final state of reaction (13). The analysis of the experimental cross sections suggests the existence of a bound state of the $\Lambda\bar{\Lambda}$ pair with a binding energy $\varepsilon_{\Lambda\bar{\Lambda}}=(36\pm 5)$ MeV, which should manifest itself in the elastic scattering cross section $e^+ + e^- \rightarrow e^+ + e^-$ in the form of a Feshbach resonance below the $\Lambda\bar{\Lambda}$ production threshold by value $\varepsilon_{\Lambda\bar{\Lambda}}$ in energy. It is also appropriate to emphasize the possibility of extracting, within the frame of the model, the low-energy scattering parameters $a_s$ and $r_0$ for short-lived hyperon pairs produced in $e^-e^+$ annihilation, which cannot be extracted from direct collisional reactions of the type (13). The paper  estimates the scattering length $\Lambda$ over $\bar{\Lambda}$ $a_s=(2.2\pm 0.3)$ fm and the effective radius of the $\Lambda\bar{\Lambda}$ interaction $r_{0} = 0.8\pm 0.4$ fm.

The effect of wave-like enhancement of the quantity $\sigma_0(E)v$ as $E\rightarrow 0$ was found by us in \cite{Mel} in analysing fusion reactions (6) and (7). In the present work, we have investigated its manifestation in the experimental cross sections of $\Lambda\bar{\Lambda}$ production at $e^-e^+$ annihilation; however, in our opinion, it should be taken into account in any two-particle near-threshold reaction. It seems to us that a natural further development of the approach would be its generalization to describe the reaction of $\Lambda_c\bar{\Lambda}_c$ pair production at $e^-e^+$ annihilation \cite{Bes3}. However, it is necessary to include the Coulomb interaction between $\Lambda_c$ and $\bar{\Lambda}_c$ in the computational scheme, as well as to take into account the d-wave (3,4). The obtained formulas could be useful for analyzing the oscillatory behavior of the electromagnetic form factors of hyperons \cite{Heid2,Heid3} and nucleons \cite{Tomasi,Yang} extracted from $e^-e^+$ annihilation reactions. Finally, the developed approach can also be useful for analyzing fusion reactions at low energies.

The author is grateful to A.B. Arbuzov, M. Hnati\v{c}, S.N. Ershov, M.A. Ivanov, L.P. Kaptari, V.A. Korobov, A.V. Kotikov, O.V. Teryaev, and Xiaorong Zhou for useful discussions and advice.



\begin{thebibliography}{40}

\bibitem{Mel} V.S.~Melezhik, Resonance amplification of the nuclear reaction X(a,b)Y near the a+X channel threshold, Nucl. Phys. A{\bf 550}, 223 (1992).

\bibitem{Bogd1} L.N.~Bogdanova, V.E.~Markushin, V.S.~Melezhik, and L.I.~Ponomarev, Nuclear fusion reaction in the mesic molecule dt$\mu$, Sov. J. Nucl. Phys. {\bf 34}, 662 (1981).

\bibitem{Bogd2} L.N.~Bogdanova, V.E.~Markushin, V.S.~Melezhik, and L.I.~Ponomarev, The nuclear fusion in the muonic molecule dd$\mu$ and the charge symmetry violation in the low energy dd interaction, Phys. Lett. B {\bf 115}, 171 (1982).

\bibitem{Taylor} J.R.~Taylor, {\it Scattering Theory: The Quantum Theory of Nonrelativistic Collisions} (Dover, New York, 1972).

\bibitem{Deser} S.~Deser,  M.L.~Goldberger, K.~Baumann, and W.~Tirring, Energy Level Displacements in Pi-Mesonic Atoms, Phys. Rev. {\bf 96}, 774 (1954).

\bibitem{Leng} K.R.~Leng, {\it Astrophysical formulae}, Vol.II (Springer, Berlin, 1974).

\bibitem{Ponomarev} L.I. Ponomarev, Muon catalysed fusion, Contemp. Phys. {\bf 31}, 219 (1990).

\bibitem{Petitjean} C.~Petitjean, P.~Ackerbauer, D.V.~Balin et al. New prescision measurements of d$\mu$d fusion, Hyperfine Int. {\bf 101}, 1 (1996).

\bibitem{Angulo} C.~Angulo, M.~Arnould, M.~Rayet et al. A compilation of charged-particle induced thermonuclear reaction rates,  Nucl. Phys. A {\bf 656}, 3 (1999).

\bibitem{Kvits} A.~A. Kvitsinsky, Chi-Yu Hu, and J.~S. Cohen, Faddeev calculations of muonic-atom collisions: Scattering and fusion in flight, Phys. Rev. A {\bf 53}, 255 (1996).

\bibitem{Bes1} M. Ablikim, M.~N. Achasov, P. Adlarson et al. (BESIII Collaboration), Measurement of the $e^+e^- \rightarrow \Lambda\bar{\Lambda}$ cross section from threshold to 3.00 GeV using events with initial-state radiation, Phys. Rev. D {\bf 107}, 072005 (2023).

\bibitem{Bes2} M. Ablikim, M.~N. Achasov, P. Adlarson et al. (BESIII Collaboration), Unraveling the structure of $\Lambda$ hyperons with polarized $\Lambda\bar{\Lambda}$ pairs, arXiv:2506.08072v2.

\bibitem{Mil3} A.~I. Milstein and S.~G. Salnikov, Natural Explanation of Recent Results on $e^+e^-\rightarrow \Lambda\bar{\Lambda}$, JETP Letters {\bf 117}, 905 (2023).

\bibitem{Mil4} S.~G. Salnikov and A.~I. Milstein, Near-threshold resonance in $e^+e^-\rightarrow \Lambda_c\bar{\Lambda}_c$ process, Phys. Rev. D {\bf 108}, L071505 (2023).

\bibitem{Mil2} A.~I. Milstein and S.~G. Salnikov, $N\bar{N}$ production in $e^+e^-$ annihilation near the threshold revisited, Phys. Rev. D {\bf 106}, 074012 (2022).

\bibitem{Halzen} F.~Halzen and A.D.~Martin, {\it QUARKS AND LEPTONS: An Introductory Course in Modern Particles Physics}, (John Wiley $\&$ Sons, New York 2018).

\bibitem{El-Bennich} B. El-Bennich, M. Lacombe, B. Loiseau, and S. Wycech, Paris $N\bar{N}$ potential constreined by recent antiprotonic-atom data and $\bar{n}p$ total cross sections, Phys. Rev. C{\bf 79}, 054001 (2009).

\bibitem{Zhou} D. Zhou and Rob G.E. Timmermans, Energy-dependent partial-wave analysis of all antiproton-proton scattering data below 925 Mev/c, Phys. Rev. C{\bf 86}, 044003 (2012).

\bibitem{Heid1} J. Heidenbauer, X.~W. Kang, and U.-G. Mei$\beta$ner, The electromagnetic form factors of the proton in the timelike region, Nucl. Phys. A{\bf 929}, 102 (2014).

\bibitem{Heid2} J. Heidenbauer, U.-G. Mei$\beta$ner, The electromagnetic form factors of the $\Lambda$ in the timelike region, Phys. Lett. B {\bf 761}, 456 (2016).

\bibitem{Heid3} J. Heidenbauer, U.-G. Mei$\beta$ner,and L.-Y. Dai, Hyperon electromagnetic form factors in the timelike region, Phys. Rev. D {\bf 103}, 014028 (2021).

\bibitem{Mil1} A.~I. Milstein and S.~G. Salnikov, Final-state interaction in the process $e^{+}e^{-}\rightarrow \Lambda_c\bar{\Lambda}_c$, Phys. Rev. D {\bf 105}, 074002 (2022).

\bibitem{Bes3} M. Ablikim, M.~N. Achasov, P. Adlarson et al. (BESIII Collaboration), Measurement of the Energy-Dependent Pair-Production Cross Sections and Electromagnetic Form Factors of a Charmed Barion, Phys. Rev. Lett. {\bf 131}, 191901 (2023).

\bibitem{Tomasi} A. Bianconi and E. Tomasi-Gustafsson, Periodic Interference Structures in the Timelike Proton Form Factor, Phys. Rev. Lett. {\bf 114}, 232301 (2015).

\bibitem{Yang} Qin-He Yang, Di Guo, Ling-Yun Dai, J. Haidenbauer, Xian-Wei Kang, and U.-G. Mei$\beta$ner, New insights into the oscillations of the nucleon electromagnetic form factors, Sci. Bullet. {\bf 68}, 2729 (2021).

\end{thebibliography}
\end{document}